\documentclass[12pt]{iopart}

\begin{document}

\title[Representations of Monomiality Principle with Sheffer-type Polynomials ...]
{Representations of Monomiality Principle with Sheffer-type Polynomials
and Boson Normal Ordering}
\author{P Blasiak$^{*,\diamondsuit}$, G Dattoli$^{\sharp}$, A Horzela$^\diamondsuit$ 
and K A Penson$^{*}$}

\address
{$^*$ Laboratoire de Physique Th\'eorique de la Mati\`{e}re Condens\'{e}e\\
Universit\'e Pierre et Marie Curie, CNRS UMR 7600, B.C. 121\\
Tour 24 - 2i\`{e}me \'et., 4 pl. Jussieu, F 75252 Paris Cedex 05, France}

\address
{$^\diamondsuit$ H.Niewodnicza\'nski Institute of
Nuclear Physics, Polish Academy of Sciences\\
ul. Eliasza-Radzikowskiego 152,  PL 31342 Krak\'ow, Poland}

\address
{$^{\sharp}$ ENEA, Dipartimento Innovazione, Divisione Fisica Applicata\\ 
Centro Ricerche Frascati, Via E. Fermi 45, I 00044 Frascati, Rome, Italy}

\eads{\linebreak \mailto{blasiak@lptl.jussieu.fr}, 
\mailto{dattoli@frascati.enea.it}, 
\mailto{andrzej.horzela@ifj.edu.pl}, 
\mailto{penson@lptl.jussieu.fr}}


\begin{abstract}
We construct explicit representations of the Heisenberg-Weyl algebra $[P,M]=1$ 
in terms of ladder operators acting in the space of Sheffer-type polynomials. 
Thus we establish a link between the {\it monomiality principle} 
and the {\it umbral calculus}. We use certain operator identities 
which allow one to evaluate explicitly special boson matrix elements 
between the coherent states. This yields a general demonstration 
of boson normal ordering of operator functions linear in either 
creation or annihilation operators. We indicate possible applications 
of these methods in other fields.
\end{abstract}

\section{Introduction}

The Heisenberg-Weyl algebra satisfying the commutation relation
\begin{eqnarray}\label{HW}
[P,M]=1
\end{eqnarray}
has attracted attention of physicists and mathematicians for a long time. 
In Quantum Mechanics the commutator of momentum and position operators provides 
the most famous example of Eq.(\ref{HW}). Also, in the second quantization 
method the boson creation $a^\dag$ and annihilation $a$ 
operators satisfy $[a,a^\dag]=1$.

Among the infinity of possible representations of Eq.(\ref{HW}) 
the simplest one is in terms of $X$, the multiplication by $x$ 
and the derivative $D\equiv\frac{d}{dx}$, as they satisfy
\begin{eqnarray}
[D,X]=1.
\end{eqnarray}
Evidently $X$ and $D$ acting on the monomials $x^n$ give
\begin{eqnarray}\label{XD}
\begin{array}{l}
Xx^n=x^{n+1},\\
Dx^n=nx^{n-1}.
\end{array}
\end{eqnarray}
In this note we shall be concerned with constructing and studying the 
representations of Eq.(\ref{HW}) in terms of the operators $M(X,D)$ and $P(D)$ 
such that the action of $M$ and $P$ on certain polynomials $s_n(x)$ is analogous to 
the action of $X$ and $D$ on monomials. 
More specifically we shall search for $M(X,D)$ and $P(D)$ 
and their associated polynomials $s_n(x)$ (of degree $n$, $n=0,1,2,...$) that satisfy
\begin{eqnarray}\label{Monomiality}
\begin{array}{l}
Ms_n(x)=s_{n+1}(x),\\
Ps_n(x)=ns_{n-1}(x).
\end{array}
\end{eqnarray}
The polynomials $s_n(x)$ are then called {\it quasi-monomials} with respect to $M$ and $P$. 
These operators can be immediately recognized as raising and lowering operators 
acting on the $s_n(x)$'s. Obviously $M$ and $P$ satisfy Eq.(\ref{HW}).
Further consequence of Eq.(\ref{Monomiality}) is the eigenproperty of $MP$
\begin{eqnarray}
MPs_n(x)=ns_n(x).
\end{eqnarray}
\noindent The polynomials $s_n(x)$ are obtained through the action of $M^n$ on $s_0(x)$
\begin{eqnarray}\label{Mn}
s_n(x)=M^ns_0(x)
\end{eqnarray}
(in the following we shall always set $s_0(x)=1$), 
and consequently the exponential generating function (egf) of $s_n(x)$ is
\begin{eqnarray}\label{G}
G(\lambda,x)\equiv \sum_{n=0}^\infty s_n(x)\frac{\lambda^n}{n!}=e^{\lambda M}1.
\end{eqnarray}
Also, if we write the quasimonomial $s_n(x)$ explicitly as
\begin{eqnarray}\label{S}
s_n(x)=\sum_{k=0}^n s_{n,k}x^k,
\end{eqnarray}
then
\begin{eqnarray}\label{SX}
s_n(x)=\left[\sum_{k=0}^n s_{n,k}X^k\right]1.
\end{eqnarray}

Several types of such polynomial sequences were studied recently using this 
monomiality principle first formulated in 
\cite{Dattoli1}, and embodied in 
Eqs.(\ref{Monomiality})-(\ref{G}).
Among the polynomials encountered in Quantum Mechanics, Hermite and Laguerre
polynomials are of Sheffer-type, whereas Legendre, Jacobi and Gegenbauer polynomials are not. 

Here we shall show that if $s_n(x)$ are of Sheffer-type 
\cite{Rainville},\cite{RomanRota},\cite{Roman}
then it is always possible to find explicit representations of $M$ and $P$. 
Conversely, if  $M=M(X,D)$ and $P=P(D)$ 
then $s_n(x)$ of Eq.(\ref{Monomiality}) are necessarily of Sheffer-type. 

\section{Sheffer-type polynomials and monomiality}

The properties of Sheffer-type polynomials are naturally handled within the so called 
{\it umbral calculus} \cite{RomanRota},\cite{Roman}. Let us recall some relevant 
facts about them  with special emphasis on their ladder structure. 
Suppose we have a polynomial sequence $s_n(x)$, $n=0,1,2,...$ 
($s_n(x)$ being a polynomial of degree $n$). 
It is called of a Sheffer A-type zero \cite{Rainville} 
(which we shall call here Sheffer-type) 
if there exists a function $f(x)$ such that
\begin{eqnarray}\label{S0}
f(D)s_n(x)=ns_{n-1}(x),
\end{eqnarray}
which is the lowering operator.
This characterisation is not unique, i.e. there are a lot Sheffer-type sequences $s_n(x)$ 
satisfying Eq.(\ref{S0}) for a given $f(x)$. We can further classify them by postulating 
the existence of the associated raising operator. 
A general theorem \cite{RomanRota},\cite{Roman} states that a polynomial sequence $s_n(x)$ 
satisfying the monomiality principle Eq.(\ref{Monomiality}) 
with an operator P given as a function of the derivative operator only, is {\it uniquely} determined 
by two functions $f(x)$ and $g(x)$ such that $f(0)=0$, $f^{'}(0)\neq0$ and $g(0)\neq0$. 
The egf of $s_n(x)$ is then equal to
\begin{eqnarray}\label{egf}
\sum_{n=0}^\infty s_n(x)\frac{\lambda^n}{n!}=\frac{1}{g(f^{-1}(\lambda))}\ e^{xf^{-1}(\lambda)},
\end{eqnarray}
and their associated raising and lowering operators of Eq.(\ref{Monomiality}) are given by \cite{Roman}
\begin{eqnarray}\label{PM}
\begin{array}{l}
P=f(D),\\
M=\left[X-\frac{g'(D)}{g(D)}\right]\frac{1}{f'(D)}\ .
\end{array}
\end{eqnarray}
Observe the important fact that $X$ enters $M$ only linearly. 
Also, the order of $X$ and $D$ in $M$ matters.
The above holds true also for $f(x)$ and $g(x)$ which are formal power series.
Any pair $M$, $P$ from Eq.(\ref{PM}) automatically satisfies Eq.(\ref{HW}).
Here are some examples of so obtained representations of the Heisenberg-Weyl algebra:
\vspace{2mm}

a)\ \ \ $M(X,D)=2X-D$, \ \ \ \ $P(D)=\frac{1}{2}D$,\vspace{2mm}

\ \ \ \ \ \  $s_n(x)=H_n(x)$ - Hermite polynomials; 
$G(\lambda,x)=e^{2\lambda x-\lambda^2}$.\vspace{2mm}

b)\ \ \ $M(X,D)=-XD^2+(2X-1)D-X-1$, \ \ \ \ $P(D)=-\sum_{n=1}^\infty D^n$,\vspace{2mm}

\ \ \ \ \ \  $s_n(x)=n!L_n(x)$ - where $L_n(x)$ are  Laguerre polynomials; 
$G(\lambda,x)=\frac{1}{1-\lambda}e^{x\frac{\lambda}{\lambda - 1}}$.\vspace{2mm}

c)\ \ \ $M(X,D)=X\frac{1}{1-D}$, \ \ \ \ $P(D)=-\frac{1}{2}D^2+D$,\vspace{2mm}

\ \ \ \ \ \  $s_n(x)=P_n(x)$ - Bessel polynomials \cite{Grosswald}; 
$G(\lambda,x)=e^{x(1-\sqrt{1-2\lambda})}$.\vspace{2mm}

d)\ \ \ $M(X,D)=X(1+D)$, \ \ \ \ $P(D)=\ln(1+D)$,\vspace{2mm}

\ \ \ \ \ \  $s_n(x)=B_n(x)$ - the exponential (Bell) polynomials; 
$G(\lambda,x)=e^{x(e^\lambda-1)}$.\vspace{2mm}

e)\ \ \ $M(X,D)=Xe^{-D}$, \ \ \ \ $P(D)=e^D-1$,\vspace{2mm}

\ \ \ \ \ \  $s_n(x)=\frac{\Gamma (x+1)}{\Gamma(x+1-n)}$ - 
the lower factorial polynomials \cite{Turbiner};
$G(\lambda,x)=e^{x\ln (1+\lambda )}$.\vspace{2mm}

f)\ \ \ $M(X,D)=(X-\tan (D))\cos^2(D)$, \ \ \ \ $P(D)=\tan (D)$,\vspace{2mm}

\ \ \ \ \ \  $s_n(x)=R_n(x)$ - Hahn polynomials \cite{Bender}; 
$G(\lambda,x)=\frac{1}{\sqrt{1+\lambda^2}}e^{x\arctan(\lambda)}$.\vspace{2mm}

g)\ \ \ $M(X,D)=X\frac{1+W_L(D)}{W_L(D)}D$, \ \ \ \ $P(D)=W_L (D)$,\vspace{2mm} 

\ \ \ \ \ \  where $W_L(x)$ is the Lambert $W$ function \cite{Knuth};\vspace{2mm}

\ \ \ \ \ \  $s_n(x)=I_n(x)$ - the idempotent polynomials \cite{Comtet}; 
$G(\lambda,x)=e^{x\lambda e^\lambda}$.\vspace{2mm}

h) The monomiality principle has also been applied to the polynomial sequences 
of more than one variable \cite{Dattoli1},\cite{Dattoli2}-\cite{Dattoli6}. 
A case in point are Hermite - Kamp\'e de F\'eriet polynomials of index $m=1,2,...$, 
of two variables $x,y$, defined as:
\begin{eqnarray}\label{Hxy}
H_n^{(m)}(x,y)=n!\sum_{r=0}^{[n/m]}\frac{x^{n-mr}y^r}{(n-mr)!r!}
\end{eqnarray}
for which (here $D_x=\frac{d}{dx}$)
\begin{eqnarray}\label{bessel}
P=D_x,\ \ \ \ \ M=x+myD_x^{m-1}.
\end{eqnarray}
Their egf is a two-variable generalization of $G(\lambda,x)$ of example a) and it reads
\begin{eqnarray}
G(\lambda,x,y)\equiv \sum_{n=0}^\infty H_n^{(m)}(x,y)\frac{\lambda^n}{n!}=e^{x\lambda+y\lambda^m}.
\end{eqnarray}
For other examples see e.g. \cite{Dattoli3}.

\section{Boson normal order and monomiality}

Now we give an application of the above formalism 
exploiting the analogy with the second quantization approach. 
We shall consider boson creation $a^\dag$ and annihilation $a$ operators 
satisfying $[a,a^\dag]=1$. 
There are two sets of states of great importance in that representation. 
First, there are number states $|n\rangle$, ($n=0,1,2,...$), which are eigenstates of the 
number operator, $a^\dag a|n\rangle=n|n\rangle$, $\langle n|n^{'}\rangle=\delta_{n,n^{'}}$. 
The action of boson operators in the number states basis is 
$a^\dag|n\rangle=\sqrt{n+1}|n+1\rangle$ and $a|n\rangle=\sqrt{n}|n-1\rangle$. 
The second set are coherent states 
$|z\rangle=e^{-|z|^2/2}\sum_{n=0}^\infty \frac{z^n}{\sqrt{n!}}|n\rangle$,
the eigenstates of the annihilation operator $a$, 
$a|z\rangle=z|z\rangle$ \cite{Klauder}. Here we will show how to use the above formulas
to calculate some special matrix elements of operator functions,
which are linear in either $a$ or $a^\dag$. As a byproduct we shall produce relations 
for the normally ordered form of these operator functions.

\noindent To make the analogy with the monomiality property
Eqs.(\ref{XD}) and (\ref{Monomiality}), it is convenient to redefine 
the number states 
\begin{eqnarray}
\widetilde{|n\rangle}=\sqrt{n!}|n\rangle
\end{eqnarray}
(note that $\widetilde{|0\rangle}\equiv|0\rangle$).
Then the creation and annihilation operators act as
\begin{eqnarray}
a^\dag\widetilde{|n\rangle}=\widetilde{|n+1\rangle}\\
a \widetilde{|n\rangle}=n\widetilde{|n-1\rangle}
\end{eqnarray}
which allows us now to make the correspondence:
\begin{eqnarray}
\begin{array}{ccl}
X&\ \ \longleftrightarrow \ \ &a^\dag\\
D&\ \ \longleftrightarrow \ \ &a \\
x^n&\ \ \longleftrightarrow \ \ &\widetilde{|n\rangle}\ ,\ \ \ \ \ n=0,1,2,...\ \ . \\
\end{array}
\end{eqnarray}
Then the operators $M$ and $P$ take the form (see Eq.(\ref{PM}))
\begin{eqnarray}\label{PMa}
\begin{array}{l}
P(a)=f(a),\\
M(a,a^\dag)=\left[a^\dag-\frac{g'(a)}{g(a)}\right]\frac{1}{f'(a)}\ .
\end{array}
\end{eqnarray}
Recalling Eqs.(\ref{Mn}),(\ref{S}) and (\ref{SX}) we get
\begin{eqnarray}
\left[M(a,a^\dag)\right]^n|0\rangle=\sum_{k=0}^n s_{n,k}(a^\dag)^k|0\rangle.
\end{eqnarray}
In the coherent states representation it yields
\begin{eqnarray}\label{M0}
\langle z|\left[M(a,a^\dag)\right]^n|0\rangle=s_n(z^*)\langle z|0\rangle.
\end{eqnarray}
Exponentiating $M(a,a^\dag)$ and using Eq.(\ref{egf}) we obtain
\begin{eqnarray}\label{EM0}
\langle z|e^{\lambda M(a,a^\dag)}|0\rangle=\frac{1}{g(f^{-1}(\lambda))}\ e^{z^*f^{-1}(\lambda)}\langle z|0\rangle.
\end{eqnarray}
By the same token one obtains closed form expressions for the following matrix elements 
( $|l\rangle$ is the $l$-th number state, $l=0,1,2,...$):
\begin{eqnarray}
\langle z|\left[M(a,a^\dag)\right]^n|l\rangle=\frac{1}{\sqrt{l!}}s_{n+l}(z^*)\langle z|0\rangle,
\end{eqnarray}
and
\begin{eqnarray}
\langle z|e^{\lambda M(a,a^\dag)}|l\rangle=\frac{1}{\sqrt{l!}}
\frac{d^l}{d\lambda^l}\left[\frac{1}{g(f^{-1}(\lambda))}\ e^{z^*f^{-1}(\lambda)}\right]\langle z|0\rangle.
\end{eqnarray}
The result of Eq.(\ref{EM0}) can be further extended to a general matrix element $\langle z|e^{\lambda M(a,a^\dag)}|z'\rangle$.
To this end recall \cite{Klauder} that $|z\rangle=e^{-|z|^2/2}e^{za^\dag}|0\rangle$ and write
\begin{eqnarray}\nonumber
\langle z|e^{\lambda M(a,a^\dag)}|z'\rangle&=&e^{-|z'|^2/2}\langle z|e^{\lambda M(a,a^\dag)}e^{z'a^\dag}|0\rangle\\\nonumber
&=&e^{-|z'|^2/2}\langle z|e^{z'a^\dag}e^{-z'a^\dag}e^{\lambda M(a,a^\dag)}e^{z'a^\dag}|0\rangle
\\\nonumber &=&e^{z^*z'-|z'|^2/2}\langle z|e^{-z'a^\dag}e^{\lambda M(a,a^\dag)}e^{z'a^\dag}|0\rangle.
\end{eqnarray}
Next, using the property $e^{-z'a^\dag}M(a,a^\dag)e^{z'a^\dag}=M(a+z',a^\dag)$ \cite{Louisell} 
we arrive at
\begin{eqnarray}\nonumber
\langle z|e^{\lambda M(a,a^\dag)}|z'\rangle&=&e^{z^*z'-|z'|^2/2}\langle z|e^{\lambda M(a+z',a^\dag)}|0\rangle\\\nonumber
&=&e^{z^*z'-|z'|^2/2}\langle z|e^{\lambda\left(a^\dag-\frac{g'(a+z')}{g(a+z')}\right)
\frac{1}{f'(a+z')}}|0\rangle.
\end{eqnarray}
Now we are almost ready to apply Eq.(\ref{EM0}) to evaluate the matrix element 
on the r.h.s. of the above equation. Before doing so we have to 
appropriately redefine the functions  
$f(x)\to \tilde{f}(x)=f(x+z')-f(z')$ and $g(x)\to \tilde{g}(x)=g(x+z')/g(z')$.
Then $\tilde{f}(0)=0$ and $\tilde{g}(0)=1$ as required by Sheffer property 
for $\tilde{f}$ and $\tilde{g}$.
(Note that these conditions are {\it not} fulfilled by $f(x+z')$ and $g(x+z')$.)
We can now write
\begin{eqnarray}\nonumber
\langle z|e^{\lambda\left(a^\dag-\frac{g'(a+z')}{g(a+z')}\right)
\frac{1}{f'(a+z')}}|0\rangle&=&\langle z|e^{\lambda\left(a^\dag-\frac{\tilde{g}'(a)}{\tilde{g}(a)}\right)
\frac{1}{\tilde{f}'(a)}}|0\rangle\\\nonumber
&=&\frac{1}{\tilde{g}(\tilde{f}^{-1}(\lambda))}\ e^{z^*\tilde{f}^{-1}(\lambda)}\langle z|0\rangle.
\end{eqnarray}
By going back to the initial functions $f$ and $g$ this readily gives the final result
\begin{eqnarray}\label{Z}
\langle z|e^{\lambda M(a,a^\dag)}|z'\rangle
=\frac{g(z')}{g(f^{-1}(\lambda+f(z')))}e^{z^*[f^{-1}(\lambda+f(z'))-z']}\langle z|z'\rangle,
\end{eqnarray}
where $\langle z|z'\rangle=e^{z^*z'-\frac{1}{2}|z^{'}|^2-\frac{1}{2}|z|^2}$ 
is the coherent states overlapping factor.

It is natural to make contact now with our solution of the 
normal ordering problem \cite{BDHPS}.
For a general function $F(a,a^{\dag})$ its normally ordered form
${\cal N}\left[F(a,a^{\dag})\right]\equiv F(a,a^{\dag})$ is
obtained by moving all the annihilation operators $a$ to the
right, using the commutation relations. We may additionally define
the operation  $:\!G(a,a^{\dag})\!:$ which means normally order
$G(a,a^{\dag})$ {\it without} taking into account the commutation
relations. Using the latter operation
 the normal ordering problem is solved for $F(a,a^{\dag})$ if we
 are able to find an operator $G(a,a^{\dag})$ for which 
$F(a,a^{\dag}) = :\!G(a,a^{\dag})\!:$ is satisfied.

\noindent To obtain normally ordered form of $e^{\lambda M(a,a^\dag)}$ we apply the 
crucial property of the coherent state representation
(see \cite {Klauder} and \cite {Louisell}). This is that if
for an  arbitrary operator ${F}(a,a^{\dag})$ we have
\begin{eqnarray}
\langle z|{F}(a,a^{\dag})|z'\rangle = \langle z|z'\rangle\ G(z^*,z')
\end{eqnarray}
\noindent then the normally ordered form of ${F}(a,a^{\dag})$ is given by
\begin{eqnarray}\label{N}
{\cal N}\left[{F}(a,a^{\dag})\right] =\ :G(a^{\dag},a):\, .
\end{eqnarray}
Eqs.(\ref{Z}) and (\ref{N}) then provide the central result
\begin{eqnarray}\label{Normal}
{\cal N}\left[e^{\lambda M(a,a^\dag)}\right]=
\ :e^{a^\dag[f^{-1}(\lambda+f(a))-a]}\frac{g(a)}{g(f^{-1}(\lambda+f(a)))}:\ .
\end{eqnarray}
Let us point out again that $a^\dag$ appears linearly in $M(a,a^\dag)$, see Eq.(\ref{PMa}). 
By hermitian conjugation of Eq.(\ref{Normal}) we obtain the expression 
for the normal form of $e^{\lambda M^\dag(a,a^\dag)}$, where $M^\dag(a,a^\dag)$ is linear in $a$.
In this context we refer to our previous work \cite{BDHPS} where the results equivalent 
to Eqs.(\ref{Z}) and (\ref{Normal}) were obtained using the substitution group approach \cite{Dattoli6}.

\noindent We shall conclude by enumerating some examples of evaluation of coherent state
matrix elements of Eqs.(\ref{M0}) and (\ref{Z}). 
We choose the $M(a,a^\dag)$'s as in the list a) -  g) above:\vspace{2mm}

a)\ \ \ $\langle z|(-a+2a^\dag)^n|0\rangle =H_n(z^*)\langle z|0\rangle$,\vspace{2mm}

\ \ \ \ \ \  $\langle z|e^{\lambda (-a+2a^\dag)}|z'\rangle=e^{\lambda(2 z^*- z')-\lambda^2}\langle z|z'\rangle$.\vspace{2mm}

b)\ \ \ $\langle z|\left[-a^\dag a+(2a^\dag-1)a-a^\dag+1\right]^n|0\rangle =n!L_{n-1}(z^*)\langle z|0\rangle$,\vspace{2mm}

\ \ \ \ \ \  $\langle z|e^{\lambda \left[-a^\dag a+(2a^\dag-1)a-a^\dag+1\right]}|z'\rangle
=\frac{1}{1-\lambda(z'-1) }e^{z^*\lambda\frac{(1-z')^2}{\lambda(1-z')-1}} \langle z|z'\rangle$.\vspace{2mm}

c)\ \ \ $\langle z|\left(a^\dag\frac{1}{1-a}\right)^n|0\rangle =P_n(z^*)\langle z|0\rangle$,\vspace{2mm}

\ \ \ \ \ \  $\langle z|e^{\lambda \left(a^\dag\frac{1}{1-a}\right)}|z'\rangle
=e^{z^*[1-\sqrt{1-2(\lambda+z'-\frac{1}{2}z'^2)}-z']}\langle z|z'\rangle$.\vspace{2mm}

d)\ \ \ $\langle z|(a^\dag a+a^\dag)^n|0\rangle =B_n(z^*)\langle z|0\rangle$,\vspace{2mm}

\ \ \ \ \ \  $\langle z|e^{\lambda (a^\dag a+a^\dag)}|z'\rangle=e^{z^*(z'+1)(e^\lambda-1)}\langle z|z'\rangle$.\vspace{2mm}

e)\ \ \ $\langle z|(a^\dag e^{-a})^n|0\rangle =\frac{\Gamma (z^*+1)}{\Gamma(z^*+1-n)}\langle z|0\rangle$,\vspace{2mm}

\ \ \ \ \ \  $\langle z|e^{\lambda (a^\dag e^{-a})}|z'\rangle=e^{z^*[\ln(e^{z'}+\lambda)-z']}\langle z|z'\rangle$.\vspace{2mm}

f)\ \ \ $\langle z|\left[(a^\dag-\tan (a))\cos^2(a)\right]^n|0\rangle =R_n(z^*)\langle z|0\rangle$,\vspace{2mm}

\ \ \ \ \ \  $\langle z|e^{\lambda (a^\dag-\tan (a))\cos^2(a)}|z'\rangle
=\frac{\cos[\arctan (\lambda+\tan(z'))]}{cos(z')}e^{z^*[\arctan(\lambda\tan(z'))-z']}\langle z|z'\rangle$.\vspace{2mm}

g)\ \ \ $\langle z|\left[a^\dag\frac{1+W_L(a)}{W_L(a)}a\right]^n|0\rangle =I_n(z^*)\langle z|0\rangle$,\vspace{2mm}

\ \ \ \ \ \  $\langle z|e^{\lambda a^\dag\frac{1+W_L(a)}{W_L(a)}a}|z'\rangle=e^{z^*[\lambda e^{\lambda+W_L(z')}+z'(e^\lambda-1)]}\langle z|z'\rangle$.\vspace{2mm}

\section{Conclusions and outlook}

In this work we have used the prescription from the umbral calculus to find 
explicit forms of the raising and lowering operators acting in the 
space of Sheffer-type polynomials. Their specific form given by Eq.(\ref{PM})
permitted us to calculate certain coherent state matrix elements.
Those in turn have given directly, see Eq.(\ref{Normal}), 
an explicit expression for the normally ordered form of operator functions of 
$M(a,a^\dag)$ since $M$ is linear in either $a$ or $a^\dag$. 
A legitimate question is if other types of sequences would be quasimonomials 
with respect to some operators $M$ and $P$. The affirmative answer was given in 
\cite{Cheikh} where it has been demonstrated that for any sequence such 
operators can be found. For a number of polynomial sequences the operators $M$ and $P$
can actually be written down explicitly. However, only in the Sheffer case $P=P(D)$ 
and $M(X,D)$ is linear in $X$. In any other case $P=P(X,D)$ and and it 
appears that this circumstance renders the calculation of coherent state 
matrix elements rather difficult. This is without doubt another demonstration
of an intrinsic difficulty to perform a normal ordering of arbitrary 
operator function.

Let us mention that the operator methods based on monomiality principle developed above
have interesting, and as yet non-explored, ramifications to at least two other fields \cite{Dattoli7}:
non-local evolution equations and generalised heat equations. 
We shall briefly sketch these points here, 
leaving the details for the forthcoming publication.
\\
Let us consider here the case c) above for which $s_n(x)$ are the Bessel polynomials
with $M(X,D)=X\frac{1}{1-D}$ and $P(D)=-\frac{1}{2}D^2+D$. 
Then it is possible to look for the solution of the following operator 
evolution equation:
\begin{eqnarray}\label{evolution}
\frac{\partial}{\partial y} F(X,y)=[P(D)+M(X,D)]F(X,y),
\end{eqnarray} 
with the 'initial' condition $F(X,0)=q(X)$.
Using $A^{-1}=\int_0^\infty e^{-sA}ds$ we find $M=X\int_0^\infty e^{-s}e^{sD}ds$,
and then with $Mq(X)=X\int_0^\infty e^{-s}q(X+s)ds$, Eq.(\ref{evolution}) takes the following 
{\em non-local} form
\begin{eqnarray}
\frac{\partial}{\partial y} F(X,y)&=&[-\frac{1}{2}D^2+D]F(X,y)+X\int_0^\infty e^{-s}F(X+s,y)ds,
\end{eqnarray} 
whose formal solution can be written as
\begin{eqnarray}
F(X,y)=e^{y(P+M)}q(X)=e^{-\frac{1}{2}y^2}e^{yP}e^{yM}q(X).
\end{eqnarray} 
For given $q(X)$, $e^{yM}q(X)$ can be evaluated recursively through 
$e^{yM}q(X)=\sum_{n=0}^\infty \pi_n(X)\frac{y^n}{n!}$ with $\pi_n(X)=X\int_0^\infty e^{-s}\pi_{n-1}(X+s)ds$.
With additional standard transformations a closed form of $F(X,y)$ can be obtained.
\\
Another perspective will be offered by further two variable \cite{Dattoli8} extension of Eq.(\ref{bessel}). 
Define
\begin{eqnarray}
\Pi=f(D_x),\\
\Theta=M_x+2M_yf(D_y)
\end{eqnarray} 
with $D_\eta =\frac{\partial }{\partial \eta }$ and 
$M_\eta =[\eta-\frac{g'(D_\eta )}{g(D_\eta)}]\frac{1}{f'(D_\eta)}$ for $\eta=x,y$.
Evidently $[\Pi,\Theta ]=1$. This particular two-variable generalisation of monomiality 
operators allows one the following extension of Hermite-Kamp\'{e} de F\'{e}riet
polynomials for $m=2$ of Eq.(\ref{Hxy}):
\begin{eqnarray}
S_n(x,y)=n!\sum_{r=0}^{[\frac{n}{2}]}\frac{s_{n-2r}(x)s_r(y)}{(n-2r)!r!}
\end{eqnarray}
which are umbral extensions \cite{Dattoli8} of the ordinary case, 
where $s_n(\eta)$ are Sheffer polynomials associated with $f$ and $g$.
It is worth noting that $S_n(x,y)$ are solutions of generalized heat equation
\begin{eqnarray}
f(D_y)S_n(x,y)=[f(D_x)]^2S_n(x,y).
\end{eqnarray}
These and related topics will be developed in subsequent publications.

\vspace{3mm}

\noindent We thank L. Haddad for important discussions.

\noindent PB wishes to thank the 
Polish Ministry of Scientific Research and Information Technology 
for support under Grant no: 1P03B 051 26.

\end{document}